# An H-band Vector Vortex Coronagraph for the Subaru Coronagraphic Extreme-Adaptive Optics System

J. KÜHN,[1,2] E. SERABYN,[2] J. LOZI,[3] N. JOVANOVIC,[3,4] T. CURRIE,[3] O. GUYON,[3]
T. KUDO,[3] F. MARTINACHE,[3,5] K. LIEWER,[2] G. SINGH,[2,3] M. TAMURA,[6,7,8]
D. MAWET,[2,9] J. HAGELBERG,[10,11] AND D. DEFRERE[12]

[1] Institute for Particle Physics and Astrophysics, ETH Zurich, Wolfgang-Pauli-Strasse 27, 8093 Zurich, Switzerland; jonas.kuehn@phys.ethz.ch
[2] Jet Propulsion Laboratory, California Institute of Technology, 4800 Oak Grove Drive, Pasadena, CA 91109, USA
[3] Subaru Telescope, National Astronomical Observatory of Japan, National Institutes of Natural Sciences (NINS), 650 North A'Ohoku Place, Hilo, HI 96720, USA
[4] Department of Physics and Astronomy, Macquarie University, 2109 Sydney, Australia
[5] Laboratoire Lagrange, Université Côte d'Azur, Observatoire de la Côte d'Azur, CNRS, Parc Valrose, Bât. H. Fizeau, 06108 Nice, France
[6] Department of Astronomy, Graduate School of Science, The University of Tokyo, Hongo 7-3-1, Bunkyo-ku, Tokyo, 113-0033, Japan
[7] Astrobiology Center, NINS, 2-21-1 Osawa, Mitaka, Tokyo 181-8588, Japan
[8] National Astronomical Observatory of Japan, NINS, 2-21-1 Osawa, Mitaka, Tokyo 181-8588, Japan
[9] Department of Astronomy, California Institute of Technology, Pasadena, CA 91125, USA
[10] Université Grenoble Alpes, IPAG, 38000 Grenoble, France
[11] Institute for Astronomy, University of Hawaii, 2680 Woodlawn Drive, Honolulu, HI 96822, USA
[12] Space Sciences, Technologies and Astrophysics Research (STAR) Institute, University of Liège, Allée du Six Août 19c, B-4000 Liège, Belgium

## ABSTRACT

The vector vortex is a coronagraphic imaging mode of the recently commissioned Subaru Coronagraphic Extreme-Adaptive Optics (SCExAO) platform on the 8-m Subaru Telescope. This multi-purpose high-contrast visible and near-infrared (R- to K-band) instrument is not only intended to serve as a VLT-class "planet-imager" instrument in the Northern hemisphere, but also to operate as a technology demonstration testbed ahead of the ELTs-era, with a particular emphasis on small inner-working angle (IWA) coronagraphic capabilities. The given priority to small-IWA imaging led to the early design choice to incorporate focal-plane phase-mask coronagraphs. In this context, a test H-band vector vortex liquid crystal polymer waveplate was provided to SCExAO, to allow a one-to-one comparison of different small-IWA techniques on the same telescope instrument, before considering further steps. Here we present a detailed overview of the vector vortex coronagraph, from its installation and performances on the SCExAO optical bench, to the on-sky results in the extreme AO regime, as of late 2016/early 2017. To this purpose, we also provide a few recent on-sky imaging examples, notably high-contrast ADI detection of the planetary-mass companion $\kappa$ Andromedae b, with a signal-to-noise ratio above 100 reached in less than 10 mn exposure time.

# 1. INTRODUCTION

Recent years have seen considerable advancement in the field of high-contrast direct imaging of low-mass companions from ground-based telescopes. To tackle the extremely challenging task of imaging the thermal emission of a young planetary-mass companion, which requires deep contrast ($10^{-6}$ to $10^{-9}$) at small angular separation, new instruments such as the Gemini Planet Imager (GPI) (Macintosh et al. 2012), the Spectro-Polarimetric High-Contrast Exoplanets REsearch (SPHERE) (Beuzit et al. 2008) and SCExAO (Jovanovic et al. 2015a) simultaneously rely on high-actuator counts kHz-regime second generation Adaptive Optics (AO) systems to correct for the atmospheric turbulences, as well as on internal active correction of non-common path aberrations (NCPAs) taking place along the scientific optical path downstream of the AO wavefront sensor (WFS) (Wallace et al. 2010; Martinache et al. 2014; N'Diaye et al. 2016; Martinache et al. 2016). The bulk of the stellar point-spread function (PSF) can then be attenuated by a few orders of magnitudes using various types of coronagraphs, located in intermediate pupil- (Kenworthy et al. 2007) or focal-plane(s) (Rouan et al. 2000; Mawet et al. 2010) upstream of the scientific focal plane array. In addition, to improve residual PSF subtraction and further reject fainter NCPA speckles, advanced observing strategies have to be employed, such as e.g. Angular Differential Imaging (ADI) (Marois et al. 2006), Spectral-/Dual-band Differential Imaging (SDI/DBI) (Lafrenière et al. 2007a), Coherent Differential Imaging (CDI) (Bottom et al. 2017), Polarimetric Differential Imaging (PDI) (Schmid et al. 2017), machine learning (Gomez Gonzales et al. 2016), and often a combination of those. Finally, advanced PSF subtraction post-processing algorithms, such as LOCI (Lafrenière et al. 2007b), PCA/KLIP (Amara & Quanz 2012; Soummer et al. 2012), or later derivatives/hybrids of these methods (Marois et al. 2010; Currie et al. 2012; Gomez-Gonzalez et al. 2017), are required to achieve the best PSF subtraction possible while maintaining sufficient throughput on bona-fide off-axis sources, ideally allowing to reach the background-limited regime at the smallest possible angular separation.

Even by reaching contrasts as deep as a few $10^{-6}$ beyond ~300 mas using a combination of the techniques listed above, less than a dozen or so planetary-mass companions have been uncovered through direct imaging so far (Bowler 2016). This is much lower than the number of confirmed detections for Radial Velocity (RV) and transit-based (mostly from Kepler) surveys, which are of the order of a few thousands each, indicating that direct imaging has still not reached its full scientific potential. That is unfortunate, as not only does high-contrast imaging probe a different parameter space than the RV or transit methods (massive planets at large *Kuiper*-belt scales separations) and enables to study planet formation and interaction with circumstellar disks *in-situ*, but it is also intrinsically complementary to these methods (e.g. by enabling high-resolution spectroscopy of atmospheres). Still, large direct imaging surveys in the past few years - essentially yielding non-detections - have already made it possible to derive some preliminary exoplanetary population statistics (Bowler 2016; Uyama et al. 2017), and the yield in terms of debris or protoplanetary disks science has been considerable. However, for direct imaging to access a larger population of hot thermally-emissive exoplanets (Mawet et al. 2012; Bowler 2016), there is a strong need for pushing towards ever smaller inner-working angle (IWA), as compared to the $> 3\lambda/D$ angular separation range that GPI and SPHERE were designed for (1 $\lambda/D$ ~ 40 mas at H-band on a 8-m class telescope).

The SCExAO platform of the 8-m Subaru Telescope is a 2$^{nd}$-generation AO high-contrast instrument (Jovanovic et al. 2015a) specifically built for small-IWA observations, relying on focal-plane phase mask coronagraphs, such as the vector vortex (Mawet et al. 2010) or eight-octants (Murakami et al. 2010) waveplates, or the Phase-Induced Amplitude-Apodization (PIAA) scheme and its variants (Guyon et al. 2005; Guyon et al. 2010), to reach within 1 to 2 λ/D from the star in the near-infrared (J- to K-band) depending on the configuration. Additionally, interferometric non-redundant spare aperture masking (SAM) modules such as VAMPIRES and FIRST (Norris et al. 2015; Huby et al. 2012) operating in the visible can reach down to 0.5 λ/D (10 mas at 700 nm), albeit at more moderate contrast (~$10^{-3}$). Key design choices have been made, which prioritize imaging at small angular separation from day one, for example by using a Pyramid Wavefront Sensor (PyWFS) to improve AO correction close to the star, and by implementing a coronagraphic Lyot-plane Low-Order Wavefront Sensor (LLOWFS) (Singh et al. 2015). Looking forward, the SCExAO platform is also conceived as a technological testbed for small-IWA coronagraphy ahead of the ELT-era, testing new concepts such as custom optics to eliminate the central obstruction (Murakami et al 2010) and reach down to 0.9 λ/D IWA, or the use of Microwave Kinetic Inductance Detectors (MKIDS) as focal-plane arrays (Marsden et al. 2012).

When aiming at ever smaller IWA, vector vortex coronagraphy (Mawet et al. 2009) has already a proven on-sky track record (Mawet et al. 2010), with the capability to image planetary mass companions down to the diffraction limit of the telescope (Serabyn et al. 2010). This has already prompted several telescope facilities to integrate a vortex-based mid-infrared (L'-band) observing mode, among which the VLT and Keck telescopes (Mawet et al. 2013a; Serabyn et al. 2017). The present work provides an overview of the vector vortex coronagraph high-contrast imaging mode of SCExAO, as of early 2017. Indeed, from roughly July 2016 onward, SCExAO has regularly achieved on-sky Strehl ratios (SR) beyond 80%, and even SR ~ 0.9 on the course of recent 2017 observing runs (Currie et al. 2017, in preparation), and this on bright stars in "good" conditions (seeing better than ~0''.5). This corresponds to reaching the so-called "extreme AO" regime where phase-mask coronagraphs such as the vortex can exponentially converge to their design null depth for unaberrated on-axis starlight. The paper is structured as follows: first, a brief reminder of the vector vortex technology is provided, with further details in the Appendix section. Second, pupil stop design considerations are exposed, and expected vortex raw contrast performances for the Subaru/SCExAO optical configuration are numerically-simulated. Third, the SCExAO near-infrared coronagraphic beam train is described, followed by the measured null depth and throughput performance using the SCExAO internal calibration source. Finally, a few examples of on-sky results during recent observing runs are presented, including raw coronagraphic images as well as ADI-only reduced datasets, for "median observing conditions" on Maunakea and total exposure times on target of about 600 s. As of 2017, the SCExAO H-band vortex coronagraphic mode is now open for "shared-risk" observations in conjunction with the HiCIAO imager (now legacy, recently decommissioned), or the CHARIS integral field spectrograph (Groff et al. 2015) in high-resolution H-band mode, and operates in the "residual tip/tilt jitter-limited" regime. The material presented is thus also intended to support prospective observers, as it also provides the tools to estimate performance improvements in the near-future.

## 2. THE SCEXAO VORTEX CORONAGRAPH: PRINCIPLE, SIMULATIONS AND INSTRUMENTAL PERFORMANCES

**2.1 Basic concepts of vector vortex coronagraphy**

As originally proposed, the vector vortex coronagraph uses the geometrical phase offset - also sometimes called "Pancharatnam phase" - introduced by manipulating the transverse polarization states of incoming light, to create a helicoidal phase plate with a singularity (the "vortex") at the center of the focal plane (Jenkins 2008; Mawet et al. 2009). In practice, as shown in Mawet et al. 2009, one can generate an optical vortex of topographic charge $n$ (= the number of helix phase jumps/wraps per revolution) using a half-waveplate design with a fast axis rotating $n/2$ times about a revolution. These can be manufactured with the liquid crystal polymer (LCP) technology, using photo-alignment to orient the birefringent LCP molecules in a continuously-rotating fashion (Nersisyan et al. 2013), or by etching sub-wavelength gratings in diamond, with an angular grove geometry to achieve form-based birefringence (Delacroix et al. 2013). In both cases, the geometrical phase delay takes place for any polarization state, and is thus independent of incoming state of polarization. Although the LCP manufacturing approach is to first order achromatic, the introduced phase shift is proportional to $n_e(\lambda) - n_o(\lambda)$ (the difference between the LCP extraordinary refractive index $n_e$ and ordinary index $n_o$), and is thus still slightly dependent on the wavelength. If required, this effect can be however mitigated over a wider bandwidth, using multi-layered designs similar to commercially-available achromatic quarter- or half-waveplates.

As illustrated in Figure 1, the operation of a vortex coronagraph is based on the diffraction properties of the vortex phase singularity which, in the absence of optical aberrations and manufacturing defects, will diffract away on-axis light coming to focus at its center. This diffracted starlight can be blocked out at the next downstream pupil plane location (in the "Lyot plane"), where an opaque mask (a.k.a the "Lyot stop") can be inserted (see Fig. 1). However this diffraction process does not apply to an off-axis source, which experiences only moderate local phase modulation at the coronagraph focal plane location, and can in turn freely propagate further downstream towards the science detector. There is a transition regime, when the off-axis source gets in the vicinity of the vortex phase singularity, where it will start to experience throughput losses due to diffraction. The IWA is thus defined as the angular separation at which an off-axis source throughput is 50%; for a perfect charge-2 vortex in the absence of aberrations (i.e. unobscured entrance pupil) the inner working angle (IWA) is ~0.9 $\lambda/D$. Figure A1 in the Appendix presents plots of theoretical throughput vs. angular separation for a few vortex topographic charges, or entrance pupil configurations.

The key features of the vector vortex coronagraph (VVC) are the diffraction-limited IWA and high throughput. For angular separations beyond a few $\lambda/D$ units, the off-axis throughput is limited only by the transmissivity of the vortex waveplate, which can be optimized beyond 95% using proper anti-reflection (A/R) coatings, and by pupil geometry (transmission area) factors, such as masking tolerance margins in the post-coronagraphic Lyot pupil mask (Fig.1). Conversely, however, the small IWA capability renders the VVC very sensitive to residual tip/tilt jitter. For unobscured apertures, the leakage (aka the null depth $N$, which is the ratio of total intensity of the coronagraphic image to the total intensity of non-coronagraphic image; see

Appendix §A.2 for detailed terminology definition) of a VVC in the presence of a small off-axis offset *t* has been analytically shown (Jenkins et al. 2008, Huby et al. 2015) to be proportional to $t^n$, where *t* is in units of λ/D, and *n* is the topographic charge. This requires maintaining AO tip/tilt residuals and non-common path instrumental tip/tilt vibrations inside a few percent of λ/D, which usually requires a dedicated low-order wavefront sensing strategy along the science beam train. Using higher topographical charges *n* vortices can dramatically reduce tip/tilt sensitivity, but at the price of giving up IWA capabilities (see Appendix Fig. A1). It can be noted that the use of more aggressive *n* = 4 vortices may be required for coronagraphic imaging of the nearest giants with the ELTs in a few years from now, as a handful of them will start to be resolved.

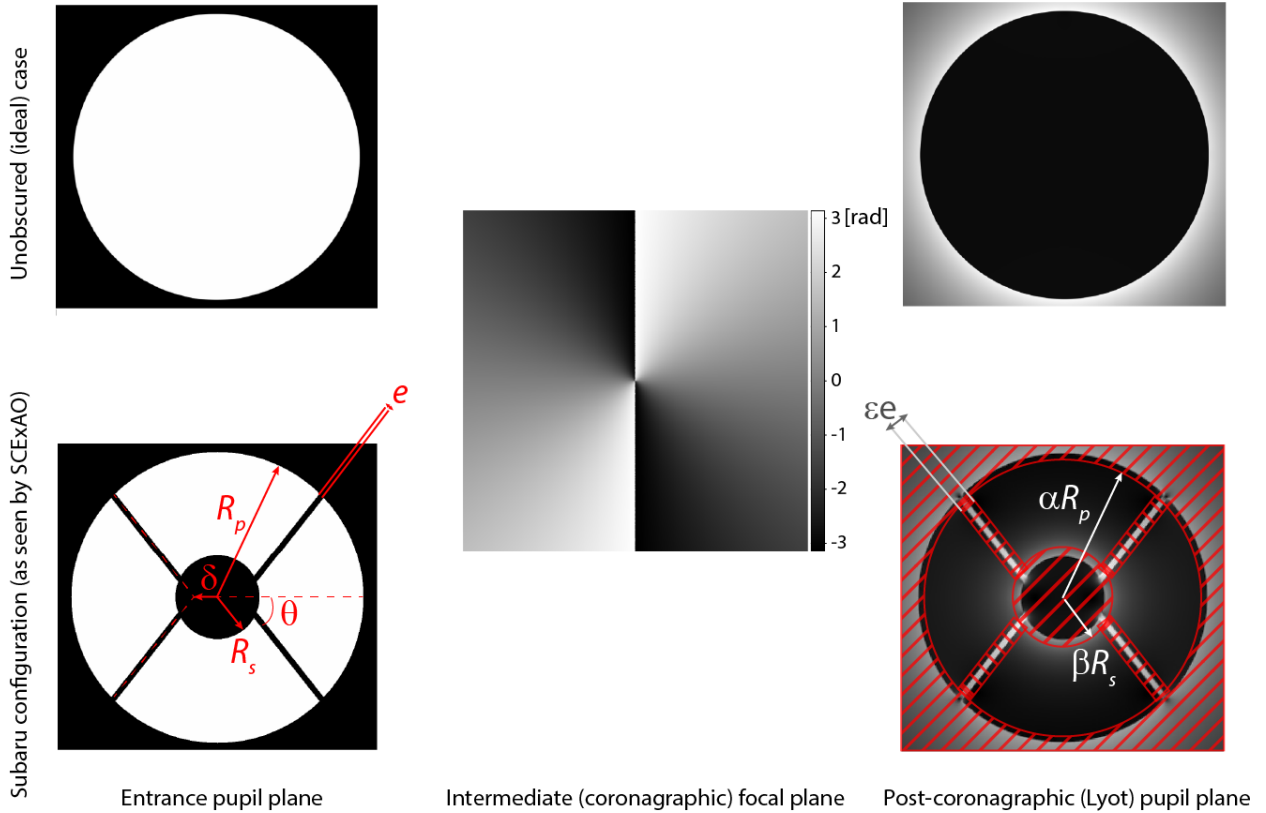

**Figure 1. Principle of vector vortex coronagraphy for ideal (upper row) and Subaru-shaped (bottom row) telescope pupils, in the case of a topographic charge-2 phase mask. The Subaru pupil and SCExAO Lyot mask parameters shown here are detailed in Section §2.3 and Table 2.**

## 2.2 A vector vortex coronagraph for SCExAO

Given that the SCExAO instrument goals are primarily focused on small-IWA coronagraphy, the integration and testing of a VVC was a natural choice, with the potential to compare performances with other focal-plane coronagraphs. SCExAO imaging path operates in the near-infrared in order to detect the thermal emission of young exoplanets, and those still undergoing formation. Because it relies on wavelength-scaled microstructures, the AGPM approach (Delacroix et al. 2013) cannot be reliably employed at wavelengths below 3.5 μm (L-band), although efforts towards manufacturing K-band AGPMs are currently ongoing. Consequently, the single-layer LCP manufacturing approach was selected for SCExAO, as already successfully implemented for the K-band vortices on the Palomar 200-inch Telescope PHARO instrument (Serabyn et al. 2010; Bottom et al. 2015). Indeed, as demonstrated in Mawet et al. 2010, the high technology readiness single-layer LCP vortices can already reach a null depth $N$ in excess of $3 \cdot 10^{-3}$ for 20% bandwidth, which is beyond the Strehl-limited regime for starlight rejection with extreme-AO systems achieving SR ~ 90-95% in the near infrared. Given that SCExAO uses a dedicated coronagraphic Lyot-plane Low-Order Wavefront Sensor (LLOWFS) (Singh et al. 2015), a charge-2 vortex waveplate design was chosen to be able to ultimately reach the best possible IWA (0.9-1.7 λ/D depending on pupil geometry, see §A.1). Finally, as the SCExAO baseline goal was to exceed 90% SR at H-band, a vortex central wavelength in H-band (1.65 μm, 10% bandwidth) was selected, to take advantage of the 40-80 mas IWA at this near-infrared (NIR) wavelength (1-2 λ/D with Subaru's 8-m aperture).

The H-band LCP vortex mask lent to SCExAO was manufactured by JDS Uniphase for the Jet Propulsion Laboratory (JPL), in Pasadena, California, using a proprietary process (McEldowney et al. 2008; Mawet et al. 2009). The substrate consists of a sandwich of fused silica, liquid crystal (ROF5104 from Rolic), and fused silica layers, with broadband near-infrared A/R coatings on both sides. Table 1 lists a few additional key mechanical and optical characteristics of this mask, among which is the "central defect" size. As detailed in Mawet et al. 2009, this corresponds to the so-called "disorientation region" at the center of the vortex singularity, where the available space is not large enough for the liquid crystal molecules to maintain the desired helicoidal alignment. It was shown that for a central defect size inferior to λ/D scales in the focal plane, chromatically-limited performance can be retained if this disorientation region is masked by an opaque metallic dot. As indicated in Table 1, the actual central defect diameter of the SCExAO vortex is ~20 μm, which was covered by a 25 μm-sized metallic dot. With the SCExAO focal ratio of F/28 on the intermediate coronagraphic focal plane (see §2.4), this scale is only about half the diffraction-limited spot size (Fλ ~ 46 μm at H-band), within the well-behaved regime described in Mawet et al. 2009.

TABLE 1

SCEXAO VECTOR VORTEX WAVEPLATE MAIN MANUFACTURING SPECIFICATIONS

| Property | Value |
| --- | --- |
| Material | Fused silica - ROF5104 liquid crystal - Fused silica |
| Vortex topographic charge | $n = 2$ |
| Design wavelength | 1.65 μm (H-band) |
| Central defect (disorientation region) size | 20 μm diameter |
| Opaque central metallic dot size | 25 μm diameter |
| Substrate size | 12.5 mm x 12.5 mm x 1.65 mm |
| Anti-reflection (A/R) coating | $T_{VVC, A/R} > 0.99$ over H-band 10% bandwidth |
| Manufacturer | JDS Uniphase |

## 2.3 Design and simulations of the SCExAO vortex coronagraphy observing mode

### 2.3.1 Working with the Subaru Telescope entrance pupil

The operational principle of the vortex coronagraph as presented above only applies for an ideal case, i.e. in absence of optical aberrations and for an unobscured entrance pupil (Fig.1, upper row). Neither condition is met for a real telescope. In particular, the Subaru entrance pupil consists of an 8.2-m primary aperture containing a 2.35-m diameter central obscuration from the secondary mirror, which is supported by four 0.23-m wide "spider" arms (see Fig.1 and Table 2). As illustrated in Figure 1, such a pupil is far from optimal for coronagraphy (Mawet et al. 2011), as it causes the secondary mirror to "leak" inside the high-contrast region of the pupil, and the support structure to "brighten". The latter phenomenon is negligible at the few $10^{-3}$ null depth level for such relative spider thickness, and this leakage can then be easily masked out in the Lyot plane, with little impact on overall throughput. However, as detailed in the Appendix Section §A.3, the secondary mirror leakage term for the general vortex case is in the order of $(R_s/R_p)^n$, where $R_p$ and $R_s$ are the primary and secondary mirror radii, respectively (Mawet et al. 2011, Serabyn et al. 2017). As an example, in the case of a topographic charge $n = 2$ vortex, a secondary to primary mirror radii ratio $R_s/R_p \sim 0.3$ would degrade the null depth to about $N \sim 10^{-1}$ in absence of any other aberrations or leakage sources (where $N \sim 0$ could otherwise be expected). Furthermore, the coronagraphic PSF then appears "donut-shaped" (Fig. 2) with stronger ringing than the theoretical Airy pattern, due to the loss of low spatial frequencies, which in turns degrades the IWA to $\sim 2\ \lambda/D$ (see Appendix §A.1).

Various solutions to tackle the central obscuration leakage from phase mask coronagraphs have been employed or proposed in the recent years. These include the use of a much reduced unobscured sub-aperture encircled between the primary and secondary (Serabyn et al. 2010), the addition of a ring amplitude apodizer in an intermediate pupil plane upstream of the coronagraph (Mawet et al. 2013b), or a two-stage vortex configuration to fold the secondary leakage flux back inside the secondary pupil location, where it can be masked out with no throughput penalty (Mawet et al. 2011). All three of these solutions would be optically invasive, and the latter two were not mature at the time of SCExAO coronagraph selection in 2012: they were only recently demonstrated on sky with the SDC instrument at Palomar (Bottom et al. 2016). Therefore we opted for the classical solution of oversizing the Lyot stop pupil mask downstream of the coronagraph, as successfully implemented on the PHARO imager at Palomar since 2012. As detailed below, in this case the design of a Lyot stop mask becomes mostly a trade-off study between throughput and null depth, with some extra allowances for mechanical positioning tolerances. Nevertheless, it should be noted that this mitigation solution to the central obscuration leakage does not prevent the IWA loss, nor does it avoid the coronagraphic PSF to become donut-shaped (Fig. 2). Only future upstream apodization optics upgrades to SCExAO coronagraphic modes, for example the Modified PIAA (MPIAA) scheme (Murakami et al. 2014), could enable the instrument to reach the $\sim 0.9\ \lambda/D$ IWA that the vortex can theoretically provide.

TABLE 2

SCEXAO ENTRANCE PUPIL AND VORTEX LYOT STOP SPECIFICATIONS

| Parameter description | Design value | Alternative metric (e.g. as projected on Subaru entrance pupil) |
|---|---|---|
| SCExAO entrance pupil (static mask between SCExAO deformable mirror and its main fore optics) | | |
| Primary mirror diameter | $D_p = 8.2$ m | $R_p = 4.1$ m |
| Ratio of secondary to primary mirror diameter | $D_s/D_p = 0.289$ | $D_s = 2.369$ m, $R_s = 1.184$ m |
| Ratio of spiders thickness to primary diameter | $e/D_p = 0.029$ | $e = 0.237$ m |
| Spiders orientation angle vs. horizontal | $\theta = 51.75°$ | |
| Spiders origin decenter along the horizontal | $\delta/D_p = 0.081$ | $\delta = 0.661$ m |
| SCExAO Vortex reflective Lyot pupil mask | | |
| Primary mirror undersize factor | $\alpha = 0.93$ | |
| Secondary mirror oversize factor | $\beta = 1.423$ | |
| Spiders masking oversize factor | $\varepsilon = 3.23$ | |
| Geometrical Lyot stop throughput | $T_{geom} = 0.7$ | |
| Lyot stop glass transmissivity | $T_{glass} = 0.91$ | $T_{total} = T_{geom} \cdot T_{glass} = 0.63$ |
| Lyot stop reflectivity for LLOWFS (see §2.4.1) | $R > 0.6$ across NIR | |
| Expected vortex null depth for SR = 1 (see §A.3) | $N = 3.8 \cdot 10^{-2}$ | $A = N^{-1} = 26$; $A_{PTP} \sim 115$ |

## 2.3.2 Simulations to derive the vortex Lyot Stop oversize factors and resulting null depth vs. Strehl-ratio performances

Devising the geometrical parameters of the post-coronagraphic pupil Lyot stop is an integral part of the design process of a focal-plane coronagraph. As detailed in Appendix §A.3, the achievable null depth of the coronagraph $N = A^{-1}$ indeed directly depends on the entrance pupil and Lyot pupil geometrical factors. The former naturally corresponds to the actual telescope pupil shape, while the latter comprise oversize or undersize factors applied on the entrance pupil features. To first order, these Lyot stop geometric parameters can be of two kinds (although there is a second-order cross-dependence): mechanical tolerances-driven or null depth-derived. As illustrated in Figure 1, for an unsegmented centrally-obscured aperture like the Subaru pupil, the Lyot stop primary mirror undersize factor $\alpha$ and the support spiders mask oversize number $\varepsilon$ will be mostly set as a function of pupil wheel positioning tolerances, combined with the Lyot mask location uncertainty in its own wheel slot. Indeed, the unwanted diffracted starlight in the vortex post-coronagraphic pupil light distribution (Fig.1) experiences a sharp nearly binary transition at these locations. As listed in Table 2, final retained values for these Lyot stop scaling factors are $\alpha = 0.93$ for the primary mirror undersize stop and $\varepsilon = 3.23$ for the spiders thickness oversize masks, essentially set with the SCExAO optical engineering team as a function of the achievable manufacturing and pupil wheel positioning tolerances (details of the study not included here).

On the other hand, the secondary mirror Lyot stop mask oversize factor $\beta$ is mostly set as a trade-off between the desired null depth $N$, as the central obstruction leakage decays as $1/r^2$ inside the Lyot pupil (see Fig.1, and Eq.6 in §A.3), and pupil throughput. An important stated goal of the SCExAO vortex coronagraphic mode was to achieve at least *100:1* peak-to-peak attenuation ($A_{ptp} > 100$, see §A.2 for terminology definition) in absence of wavefront errors (SR =1) in H-band with a bandwidth of 10%. Several numerical simulations using Fourier optics were conducted using *IDL* to study the influence of the Lyot stop geometry on the null depth, and to converge on the final choice of the secondary mask oversize factor $\beta$ (see Fig.1). As indicated in Table 2, the final value of $\beta \sim 1.4$ for the oversize factor on the Lyot stop secondary mask was chosen, blocking out about two thirds of the integrated secondary leakage term. This yields a simulated total attenuation of $A \sim 26$, i.e. a null depth of $N = 3.8 \cdot 10^{-2}$, corresponding to a peak-to-peak attenuation $A_{PTP} \sim 112$, in the absence of wavefront errors ($SR = 1$) and for a bandwidth of 10% in the H-band. Indeed, knowing the Subaru and retained Lyot stop pupil geometries, an estimate of the peak-to-peak attenuation $A_{PTP}$ can be computed as $A_{PTP} \sim 4.2 \cdot A$ (only valid for $SR \sim 1$, see §A.2 for details) owing to the "donut" reshaping of the coronagraphic PSF (see Fig.2). To first order, this means that for high Strehl ratio regime ($SR > 0.95$), the vortex coronagraph should enable to integrate more than 50 times longer before saturating on the science focal-plane array, as compared to the non-coronagraphic case.

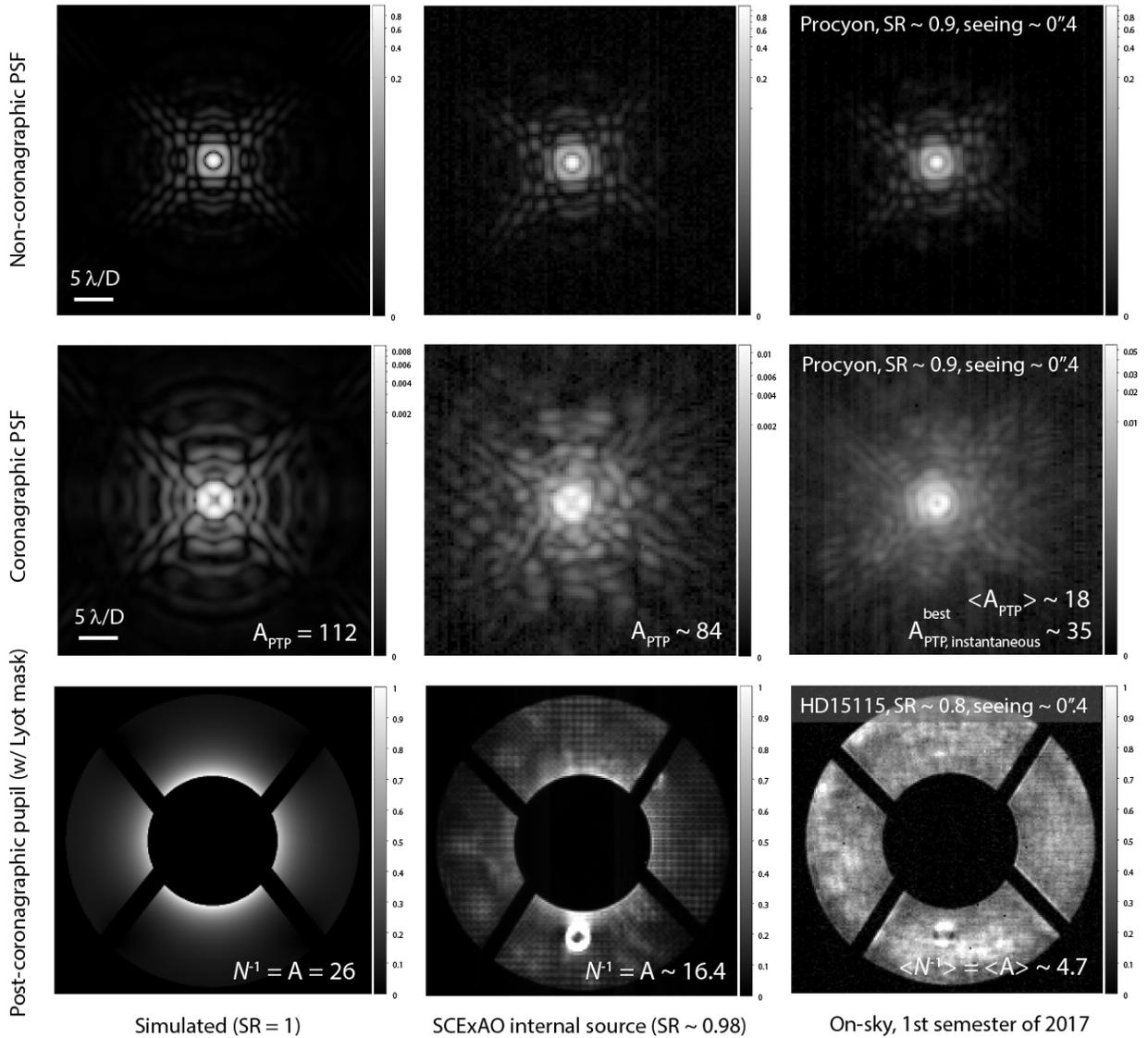

Figure 2. SCExAO vortex coronagraph numerically simulated (left), instrumental (middle) and on-sky (right, with DIMM seeing quoted) broadband point-spread functions (PSFs), and post-coronagraphic Lyot pupil planes (with the Lyot mask in place). All PSF images are in logarithmic scale normalized to the peak of non-coronagraphic PSFs.

Overall, assuming the Lyot stop geometry of Table 2, and the absence of wavefront errors (SR=1) or any other extra leakage sources (see §A.4-6), simulated theoretical non-coronagraphic and coronagraphic PSFs, as well as post-coronagraphic pupil plane light distribution with the Lyot stop in place, are shown in Figure 2. Theoretical attenuation curves in function of Strehl ratio are plotted on Figure 3, for both the monochromatic and broadband cases.

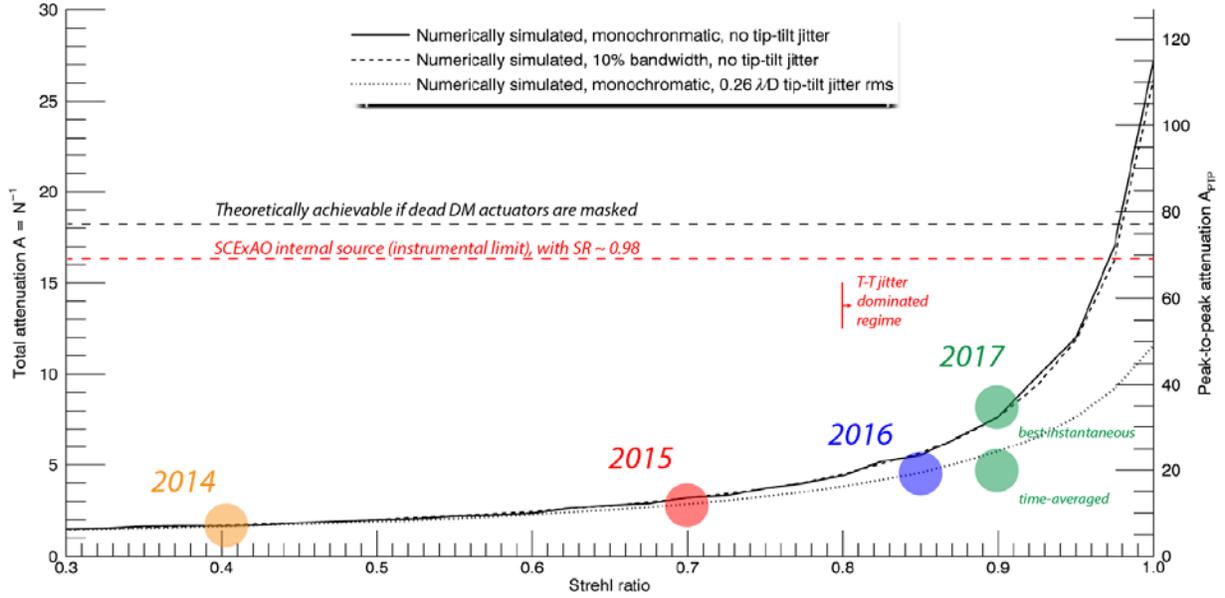

**Figure 3. Numerically simulated SCExAO vortex null depth performances as a function of Strehl ratio. Various leakage contributors (see Appendix) are added to illustrate their respective impact on null depth. Instrumental (dashed horizontal lines) and on-sky (colored dots) measurements are superimposed to show the actual performances progress over the years. The right-hand side $A_{PTP}$ vertical axis assumes a coronagraphic PSF geometric scaling factor $A_{PTP}/A \sim 4.2$, but this quickly loses validity for $SR < 1$.**

Taking into account all the Lyot stop geometrical oversize/undersize factors (see Table 2), the geometrical throughput for the SCExAO vortex Lyot pupil mask is $T_{geom} = 0.7$, which is still relatively competitive compared to ring-apodizers solutions (Mawet et al. 2013b). It is worth noting that a non-negligible throughput gain (up to 10-20%) could be obtained by reducing the margins on the primary undersize parameter $\alpha$, when SCExAO optical path will be frozen for good, and the pupil-plane wheel positioning repeatability appropriately assessed. Also, as previously mentioned, the SCExAO vortex Lyot stop is implemented as a transmissive glass substrate. Indeed, and uniquely to SCExAO, the blocked starlight is reflected back at a slight off-axis angle towards a dedicated near-infrared Lyot-based low-order wavefront sensor (LLOWFS, see §2.4), capable of sensing residual tip/tilt and other low-order aberrations with a bandwidth of about 100 Hz (Singh et al. 2015). The transmissivity of this Lyot stop is $T_{glass} = 0.9$, which, when combined with the vortex waveplate A/R coating transmissivity of $T_{VVC,\,A/R} \sim 0.99$ (see Table 1), yields a true overall throughput $T_{total} = T_{geom} \cdot T_{glass} \cdot T_{VVC,\,A/R} = 0.62$ for the vortex beam path. This number has then to be multiplied by the vortex coronagraphic throughput at a given angular separation (>95% past 8 λ/D, see Appendix §A.1), to accurately assess the overall throughput on an off-axis source. Finally, the impact of the Lyot stop geometry (mainly the central obstruction oversizing) on the IWA can be observed on Figure A1 in Annex §A.1, where it is shown that the IWA is degraded to ~1.7 λ/D.

## 2.4 Integrating the vortex coronagraph on the SCExAO instrument

### 2.4.1 Overview of the SCExAO near-infrared imaging beam train

As detailed in Jovanovic et al. 2015a, the SCExAO instrument is located on the Nasmyth platform of the Subaru Telescope, between the AO188 first generation AO system, which delivers a correction with SR ~ 30-40% at H-band (Minowa et al. 2010), and the near-IR imager HiCIAO (Hodapp et al. 2008) or CHARIS (Groff et al. 2015). SCExAO makes use of all the light from 600 to 2500 nm, splitting the 600-950 nm R-y visible and 950-2500 nm y-K NIR bands into two separated optical breadboards, mounted on top of each other. The visible portion of the spectrum is used for SCExAO wavefront sensing (unmodulated pyramid WFS), as well as by two sparse aperture masking instruments (see §1) that can operate simultaneously with NIR observing modes. Wavefront control to reach the extreme AO regime is achieved with a Boston Micromachines 2k DM (45x45 actuators across the illuminated pupil) operating at up to 2 kHz.

An optical layout of the beam train section between AO188 and the NIR imagers is provided in Figure 4, essentially a simplified unfolded version of the comprehensive layout detailed in Jovanovic et al. 2015a, as configured in vortex coronagraphy mode. Following a reflection at the 2k DM located in a pupil plane conjugated with the telescope primary mirror, the beam immediately goes through a built-in pupil mask, which stays fixed (aligned and clocked) with the entrance telescope pupil for all configurations, given that SCExAO operates in pupil-tracking mode. Then a dichroic filter sends the VIS portion of the spectrum to the visible bench (not shown here), while the NIR beam is brought into focus onto the intermediate coronagraphic focal plane at F/28, where a motorized wheel comprising several coronagraphic masks is located. A few focal-plane coronagraphic optics can be selected there, which include the PIAA and PIAA Complex Mask Coronagraph (PIAACMC) PSF blocking masks (Guyon et al. 2010), four- and eight-octants phase masks (Rouan et al. 2000; Murakami et al. 2010), a set of classical Lyot coronagraphs with mask sizes from 160 to 620 mas in diameter, and the H-band vortex coronagraph presented herein.

The diverging beam is then re-collimated by an off-axis parabolic mirror (OAP) onto a subsequent post-coronagraphic (Lyot) pupil plane, where a reflective Lyot stop is located. As previously introduced, this specific Lyot stop arrangement is uniquely available on SCExAO, and enables Lyot-plane low-order wavefront sensing (LLOWFS, see Singh et al. 2015) in-situ, using the coronagraphically-rejected starlight. Practically, the reflected starlight is focused on an InGaAs CMOS camera that serves as a low order wavefront sensor. This scheme is capable of correcting up to 35 Zernike modes at 100 Hz, and was designed to reduce tip-tilt jitter residuals to less than 1 mas on-sky (Singh et al. 2015), a critical capability for vortex coronagraphy at small IWA. Finally the beam is relayed to a beam-splitter that can send part (or all) of the light to either an internal InGaAs NIR engineering camera used for various internal alignment checks, or towards the science focal plane arrays. As indicated in Figure 4, both upstream and downstream pupil planes also integrate additional flip-able optics or wheels to configure the beam path for

PIAA coronagraphy (Guyon et al. 2010; Guyon et al. 2015) or insert additional optics in the future.

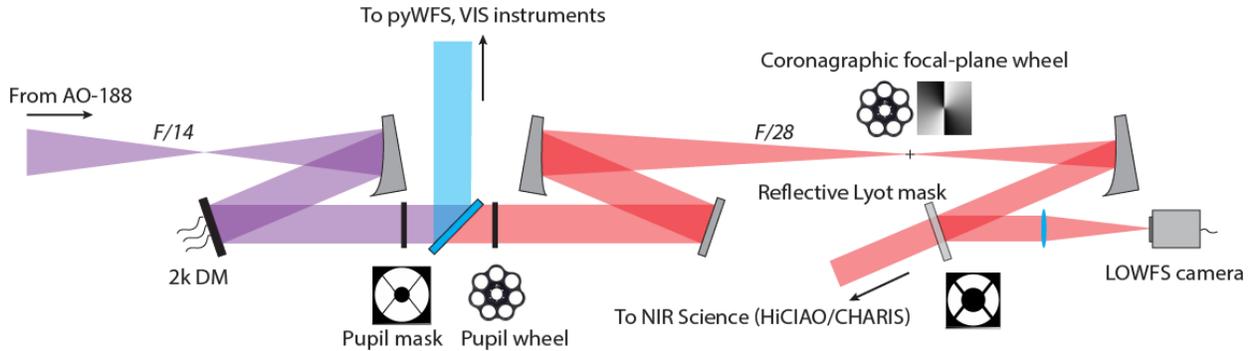

**Figure 4. Overview of SCExAO near-infrared coronagraphic optical path. This is a simplified unfolded version of the integral optical layout provided in Jovanovic et al. 2015a.**

*2.4.2 Vortex installation and instrumental performances with SCExAO internal source*

The H-band vortex coronagraph was tested at JPL during the course of 2012, and then installed inside the SCExAO bench in December 2012. As mentioned in §2.4.1, SCExAO integrates a fixed entrance pupil mask, therefore all coronagraph optics can be tested with telescope-like pupil conditions in the laboratory, using the instrument internal source (see Jovanovic et al. 2015a for details). The measured focal-plane PSFs (with/without coronagraph) and Lyot pupil plane light distribution (with coronagraph), as imaged with SCExAO internal NIR camera using the SCExAO internal source at H-band 10% bandwidth, are shown in Figure 2. In all cases, the dedicated vortex Lyot mask (Table 2) is always inserted. As indicated on Figure 2, the measured instrumental total attenuation is about *A ~ 16.4 (N ~ $6.1 \cdot 10^{-2}$)*, which – looking at the simulated curves of Figure 3 – is consistent with an instrumental Strehl ratio of *SR ~ 0.98*. It is worth mentioning the leakage due to a few dead DM actuators (~1.5 actuators), which can be observed in the Lyot pupil plane image of Figure 2. To first order, this leakage contributes to about 10% of the overall remaining light, hence masking it by manufacturing a dedicated Lyot stop (or modifying the existing one) should improve the total rejection to about 18:1. Finally, azimuthally-averaged coronagraphic and non-coronagraphic raw contrast curves corresponding to the PSF images of Figure 2, are plotted on Figure 5. The raw contrast gain past 2 $\lambda$/D (80 mas), as compared to the non-coronagraphic case, is of the order of 3 magnitudes (a factor 20), consistent with the measured total attenuation.

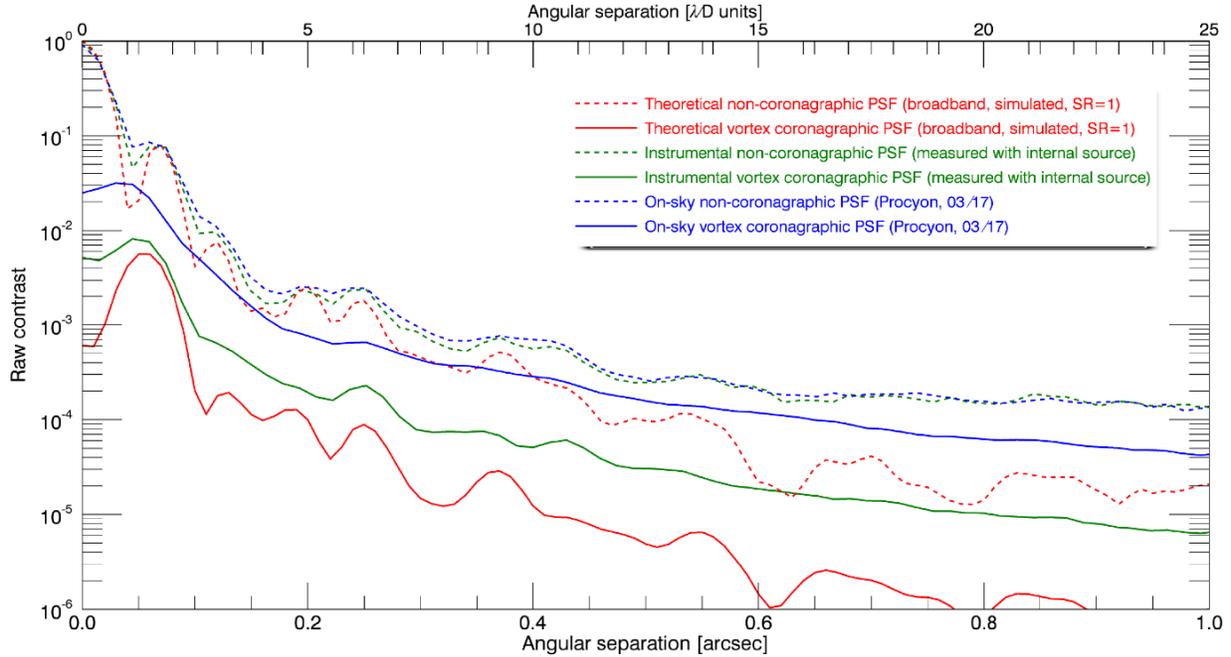

Figure 5. Raw contrast (azimuthally-averaged flux normalized to the peak of the non-coronagraphic PSF) vs. angular separation. (Red) numerically simulated, (green) measured instrumental (with SCExAO internal source), and (blue) on-sky contrast curves are plotted for both the coronagraphic and non-coronagraphic cases. All presented measurements were recorded with SCExAO internal InGaAs camera. On-sky target was *Procyon* as observed on the engineering night of March 12, 2017, UT, under ~ 0''.4 DIMM seeing.

## 3. FIRST-LIGHT AND ON-SKY COMMISSIONING RESULTS

### 3.1 Typical on-sky raw contrast performances on bright stars under 0''.4 seeing

While the vortex coronagraphic mode was undergoing commissioning in extreme AO conditions between late 2016 and early 2017, a useful engineering task was systematically undertaken at the beginning or the end of a scientific exposure sequence. These coronagraphic performance calibration sequences consisted of recording a set of unsaturated non-coronagraphic and coronagraphic PSF exposures (typically 300s per PSF stack) on bright stars (Hmag < 6), within the linear regime of the NIR camera (either HiCIAO or the internal engineering camera). Non-coronagraphic configurations have to be set with the exact same optics in the path as in the coronagraphic case, to make the procedure insensitive to each optics throughput properties: this can be achieved either by pointing the star away from the vortex center (aka tip-tilt offset, ideally at least $20\ \lambda/D$, see Fig. A1), or by offsetting the focal-plane wheel (see Fig. 5) position by the required amount of translation. Not only is such an "on-sky null depth" measurement a critical step in the commissioning path for a coronagraph (and/or the AO system), but it also provides a few useful metrics for optimal exploitation of the science exposures, namely on-sky attenuation, raw contrast curves, Strehl ratio (PSF geometry), calibration photometry, and detector-plane residual tip/tilt jitter (see below).

A subset of temporally averaged vortex null depths measured through the calibration sequences described above, obtained on very bright stars (Hmag < 4) and in good seeing conditions (seeing ~ 0''.4) since the year 2014, are shown on Figure 3 in the form of data points for each time period over the lifespan of SCExAO commissioning, overlaying the simulated curves. As seen on Figure 3, over the course of the 2014-2016 time period, the measured total attenuation numbers seemed to follow the expected behavior as a function of the on-sky Strehl ratio delivered by SCExAO through the course of its development. However, the latest null depth measurements (end 2016/early 2017) are clearly hitting a ceiling around $A_{PTP}$ ~ 20 (A ~ 5), despite clear progress on the extreme AO front over the same time period ($SR$ ~ 0.9 on bright stars like Procyon). While most of these coronagraphic calibration sequences were obtained with the science-grade HiCIAO imager over the course of the 2014-16 period, with framerates of 1 Hz or less, a recent (March 12, 2017, UT) sequence on the bright star *Procyon* was recorded with the fast framerate SCExAO internal InGaAs camera at a framerate of about 170 Hz. Further examination of the high-framerate non-coronagraphic PSF dataset (3,000 frames at 20 μs integration time) on *Procyon*, revealed a residual tip/tilt jitter at this frequency, in the order of *0.25 λ/D* tip/tilt rms (10 mas rms). This seems consistent with the "bump" observed around 200 Hz (Lozi et al. 2016), as this frequency is also outside the ~ 100 Hz bandwidth of the LLOWFS used during this on-sky calibration sequence. Poor tip/tilt jitter residuals as an explanation for the limited null depth is further reinforced by investigating the evolution of the instantaneous attenuation in time, within the high-speed *Procyon* coronagraphic time series (3,000 frames at 1 ms integration time), exhibiting instantaneous peak-to-peak attenuation numbers as high as $A_{PTP}$ ~ *35*, while the mean attenuation remains limited to $<A_{PTP}>$ ~ *18*. Both these "instantaneous best case" and time-average numbers are indicated on Figure 2, presenting the on-sky images on *Procyon*, as well as overlaid data points on Figure 3.

As detailed in Appendix §A.6, one can expect considerable coronagraphic leakage when in presence of non-negligible residual tip/tilt jitter of the order of fractions of λ/D. New *IDL* simulations were undertaken with a *0.25 λ/D* rms tip/tilt jitter residual in the coronagraphic focal-plane, and the corresponding curve for the simulated attenuation vs. Strehl ratio is over-plotted on Figure 3. Unsurprisingly, this simulation confirms that peak-to-peak attenuations of less than 25:1 can be expected even for Strehl ratio as high as *SR ~ 0.9*. At the same time, Figure 3 also shows that measured instantaneous best peak-to-peak attenuations $A_{PTP}$ ~ *40* are also typical of *SR ~ 0.9* regime free of tip/tilt jitter.

Finally, raw azimuthally-averaged "typical" on-sky raw contrast curves are plotted on Figure 5 for the latest *Procyon* observation, for both the coronagraphic and non-coronagraphic PSF data sets. The on-sky coronagraphic raw contrast performance gain is of the order of ~1.5 magnitudes (a factor 4) beyond 2 λ/D, as compared to the non-coronagraphic case, which seems consistent with the measured tip/tilt jitter-limited total attenuation. All these on-sky metrics measured during commissioning, combined with the simulated expected performances of the SCExAO vortex coronagraph as used with a Lyot stop set up as in Table 2, are summarized in Table 3. An observer who would like to plan an observation in the SCExAO-vortex-HiCIAO/CHARIS configuration should therefore refer to Table 3, which provides "typical 0''.4 DIMM seeing" performances in the presence of the ~200 Hz residual tip/tilt jitter still observed in early 2017. These on-sky performance numbers should consequently be seen as lower limits for the near-future, which will depend on progress in tackling the jitter issue.

TABLE 3

SCEXAO VORTEX HIGH-CONTRAST IMAGING MODE COMMISSIONING PERFORMANCES

| Parameter description | Commissioning value (as of 2017A) |
|---|---|
| Operating wavelength | $\lambda_0 = 1.65$ μm, 10% bandwidth |
| Inner working angle (50% coronagraphic throughput) | $IWA = 1.7\ \lambda/D$ (70 mas) |
| Coronagraphic throughput @ 3 $\lambda/D$ (120 mas) | $T_{coro} = 0.8$ |
| Coronagraphic throughput @ 5 $\lambda/D$ (200 mas) | $T_{coro} = 0.9$ |
| Overall light efficiency outside coronagraphic throughput | $T_{LS,\ geom} \cdot T_{LS,\ glass} \cdot T_{VVC,\ A/R} = 0.62$ |
| Theoretical (simulated) null depth for $SR = 1$ | $N = 3.82 \cdot 10^{-2}$; $A = 26.2$; $A_{PTP} = 111$ |
| Theoretical (simulated) null depth for $SR = 0.98$ | $N = 6.13 \cdot 10^{-2}$; $A = 16.3$; $A_{PTP} \sim 69$ |
| Theoretical (simulated) null depth for $SR = 0.9$ | $N = 1.30 \cdot 10^{-1}$; $A = 7.7$; $A_{PTP} \sim 33$ |
| Instrumental (measured) null depth with $SR_{SCExAO} \sim 0.98$ | $N = 6.10 \cdot 10^{-2}$; $A = 16.4$; $A_{PTP} \sim 84$ |
| Instrumental (measured) raw contrast @ 3 $\lambda/D$ (120 mas) | $6.5 \cdot 10^{-4}$ |
| Instrumental (measured) raw contrast @ 5 $\lambda/D$ (200 mas) | $2.2 \cdot 10^{-4}$ |
| On-sky performances on bright stars, $SR_{SCExAO} \sim 0.9$, median seeing $\sim 0''.4$, as of the 1st semester of 2017 | |
| Best instantaneous (measured) peak-to-peak attenuation | $A_{PTP,\ best} \sim 35$ |
| Time-averaged (measured) null depth with ~200 Hz residual jitter | $N = 2.13 \cdot 10^{-1}$; $A = 4.7$; $A_{PTP} \sim 18\text{-}20$ |
| Theoretical (simulated) null depth for $SR = 0.9$ and $\sigma_{TT} = 0.26\ \lambda/D$ | $N = 1.82 \cdot 10^{-1}$; $A = 5.5$; $A_{PTP} \sim 23$ |
| Typical on-sky (measured) raw contrast @ 3 $\lambda/D$ (120 mas) | $3.4 \cdot 10^{-3}$ |
| Typical on-sky (measured) raw contrast @ 5 $\lambda/D$ (200 mas) | $8.1 \cdot 10^{-4}$ |

### 3.2 Angular Differential Imaging (ADI) example: detecting *kap And b* with SNR ~ 107

The young star *κ Andromedae* (B9IVn, M ~ 2.8 M☉, 51 pc, 47 Myr), hereafter denoted *kap And*, was observed during the course of the night of November 12, 2016, UT, in order to roughly assess the raw ADI performance of the vortex coronagraph mode in the extreme AO regime, as achieved by SCExAO by late 2016. We obtained 40 15-s HiCIAO co-added frames (10 co-adds of 1.5-s individual exposures each) for a total integration time (after frame selection) of 570 s, under ~0''.4 seeing conditions. This relatively short time on target enabled to achieve 14.8 degrees of field rotation for ADI. Unsaturated PSF exposures, and various background, dark and flat-fielding sequence acquisitions were also obtained to calibrate photometry and perform cosmetics processing steps. As shown in Figure 6, the ~22 $M_J$ (Carson et al. 2013; Jones et al. 2016) companion at ~1'' is readily visible in the raw co-added 15-s frames after basic frame cosmetics (i.e. flat-fielding, background subtraction, bad pixel correction, etc…).

After frame registration using SCExAO DM-generated incoherent artificial satellite spots (Jovanovic et al. 2015b) for centroiding (shift and subtract method), we first performed a simple median-subtract ADI reduction (Marois et al. 2006) in order to provide a raw contrast curve independent of any particular choice of reduction parameters. Figure 6 presents this final median-combined ADI image after derotation, as well as a few contrast curves: the basic ADI 5-$\sigma$ contrast curve (adjusted for self-subtraction/throughput, but not for small sampling statistics [Mawet et al. 2014]), an example of Adaptive LOCI (ALOCI, see Curie et al. 2012) reduction (adjusted for throughput and small sample statistics), and the current best throughput-adjusted contrast obtained with the vortex on-sky (HD36546 on October 5, 2016 UT [Currie et al. 2017)]), with about 4 times longer integration time and 7 times larger position angle (PA) motion. At a magnitude ratio of *ΔH = 10.35* to the host star, the companion is detected with a SNR from 72 (median-combine ADI) to 107 (ALOCI), with its first Airy ring being visible. Using a ~8.3 mas/pixel platescale for HiCIAO (Currie et al. 2017), and in the absence of an astrometric field calibration frame, rough astrometry for *kap And b* obtained at this epoch yields an angular separation of 944 mas (~48 AU projected separation), and a position angle of 39.1 degrees East of North. As seen on Figure 6, a basic ADI contrast (5-sigma) of about $8 \cdot 10^{-5}$ is reached at 0''.3, and $10^{-5}$ is achieved past 0''.7 (see Table 3). Another one to two magnitudes ALOCI contrast gain can be obtained on deeper exposure ADI datasets with larger parallactic angle rotation (see e.g. Currie et al. 2017).

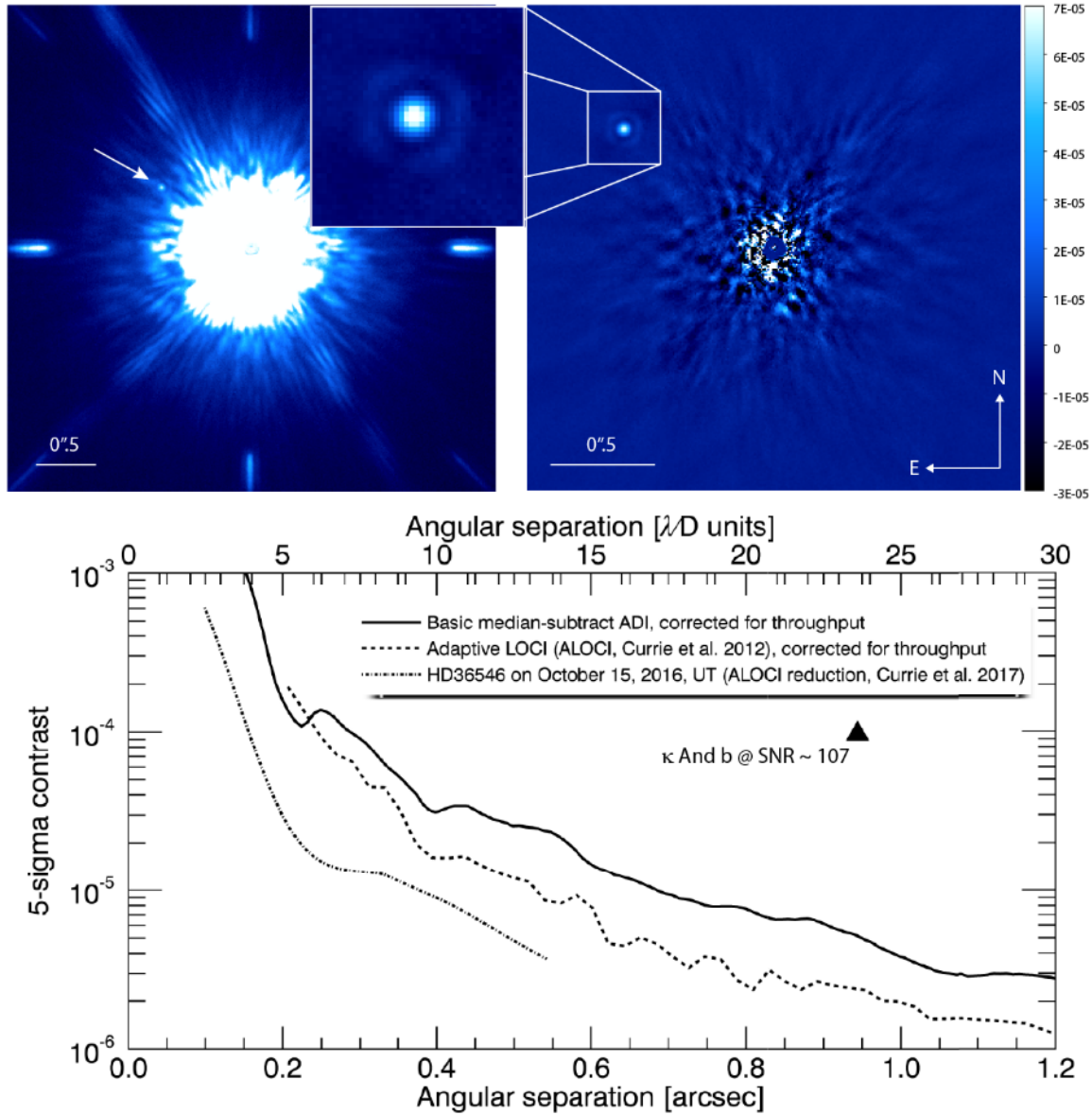

**Figure 6.** SCExAO-vortex on-sky performances on *κ And*, as obtained on the night of November 12, 2016, UT (seeing ~ 0''.4). The planetary-mass companion is visible in the raw HiCIAO co-added frames (upper left, 15 s exposure), and is clearly detected with a SNR of ~ 72 in the ADI median-combined image (upper right) and SNR ~ 107 using ALOCI. Presented reduced images are normalized by the peak photometry of the primary. The bottom panel shows 5-σ contrast curves for *κ And* (median-combine ADI and ALOCI), as well as for HD36546 on a deeper dataset, also using the ALOCI algorithm (Currie et al. 2017).

## 4. CONCLUSIONS AND PERSPECTIVES

The present work describes the challenging aspects of implementing a state-of-the-art focal-plane phase mask coronagraph, such as the vector vortex, in an extreme AO instrument downstream of a non-ideal centrally-obscured telescope pupil. Simulated, analytical and instrumental performance (with the SCExAO internal light source) results are in good agreement, and indicative of an internal Strehl ratio of ~0.98. Of course, further optimization efforts aimed at improving the coronagraphic null depth and inner working angle – for example through the use of pupil apodizing optics, or by masking the few dead DM actuators – are in the works, or could be considered. Admittedly, though, from an observer point-of-view, and as supported by the presented on-sky data, the actual high-contrast performance mostly depends on wavefront control aspects rather than from an extreme optimization of the coronagraphic stage. Indeed, while residual tip/tilt jitter is arguably the main contrast limiting factor at the moment, higher order wavefront errors still play a critical role. This is because not only does the coronagraphic null depth exponentially depend on the Strehl ratio, but NCP speckles are potentially responsible for higher false positive detection rates. This is why wavefront control is currently the main focus of the SCExAO engineering team, with a particular emphasis on low-order wavefront sensing (LOWFS), notably by improving the bandwidth of the SCExAO LLOWFS module and tip/tilt telemetry handling by the real-time system, and focal-plane wavefront sensing in general. The latter – possibly combined with coherent differential imaging (CDI) methods – will be key to reach better contrast, especially close in, where it matters the most. In this regard, the upcoming installation of the MKIDS-based integral field unit (IFU) in 2018 holds a promising potential, paving the route to a complete integration of coronagraphic and wavefront control aspects.

In the near-term, the charge-2 vector vortex coronagraph installed on SCExAO, and presented in details here, will remain available to observers in shared-risk mode. However, as the HiCIAO imager was recently decommissioned, the new baseline observing mode for the SCExAO vortex will be high-resolution spectroscopy in H-band using the CHARIS IFU. Discussions are underway regarding the potential procurement of a broadband (J-K or y-K) vortex waveplate, to be able to exploit the full potential of the IFS mode when combined with the ease-of-use and IWA of a vortex coronagraph. This study will be of course informed by the progress status on the reduction of residual tip/tilt jitter, as we note that – under particular circumstances – a topographic charge-4 vortex waveplate might be more suitable in sub-optimal observing conditions (tip/tilt jitter, but also e.g. wind gusts). Finally, it should also be stressed that an AGPM vortex operating in the mid-infrared is currently operational in NIRC2 on the neighboring Keck Telescope, which - when combined with the presented SCExAO NIR vortex capability – provides an interesting synergy in terms of wavelength of operation and high-contrast small-IWA coverage capabilities, under the Maunakea sky.

Part of this work was performed at the Jet Propulsion Laboratory (JPL), California Institute of Technology, under contract with NASA. The vortex coronagraph presented herein is a loaned part by JPL to the SCExAO instrument team. The development of SCExAO was supported by the JSPS (Grant-in-Aid for Research #23340051, #26220704 #23103002), the Astrobiology Center (ABC) of the National Institutes of Natural Sciences, Japan, the Mt Cuba Foundation and the directors contingency fund at Subaru Telescope. JK and JH are/were supported by the Swiss National Science Foundation (SNF) for this work, through the grants #PZ00P2_154800 and

APPENDIX

*A.1. Off-axis transmission*

Despite the fact that a vector vortex coronagraph relies on a central phase singularity to perfectly diffract unaberrated on-axis light outside the downstream pupil plane, the transmission behavior is obviously not binary between "block-all" and "let through". There is indeed a transition regime between on-axis and close-separation off-axis regimes, due to the local high spatial frequency derivative (or acquired orbital momentum) at close angular separations, i.e. in the vicinity of the vortex singularity. As can be expected, the higher the vortex topographical charge *n* (i.e. the number of phase jumps per 360 deg revolution), the higher the spatial frequency derivative at a given separations, hence a higher attenuation at small offsets. An accurate expression for the null depth $N = A^{-1}$ in function of small angular offset (inside ~ *0.15 λ/D*) have been derived by Huby et al. 2015 (see also Serabyn et al. 2017) in the case of an unobstructed aperture:

$$N(t) \sim \frac{\pi^2 t^2}{8} \tag{1}$$

for a charge-2 vortex, with *t* being the tip/tilt offset in units of λ/D, and

$$N(t) \sim \frac{\pi^4 t^4}{32} \tag{2}$$

in the case of a charge-4 vortex (Jenkins et al. 2008).

In Figure A1, we present a few simulated examples, using our *IDL* software code: the cases of charge-2 and charge-4 vortices for unobstructed apertures, but also for the actual SCExAO pupil configuration, using the Lyot stop configuration as in Table 2. The IWA is defined as the angular separation at which a point source experiences a transmission of 0.5, and it corresponds to *1.7 λ/D* for the charge-2 vortex currently equipping SCExAO. In Fig. A1, we can also notice a "shoulder" or "plateau" in terms of throughput for both the charge-2 and charge-4 vortex when dealing with the Subaru pupil: this can be explained by the interplay between the donut shaped PSF (stronger Airy rings) and the 25-μm sized metallic dot masking the central defect of the vortices.

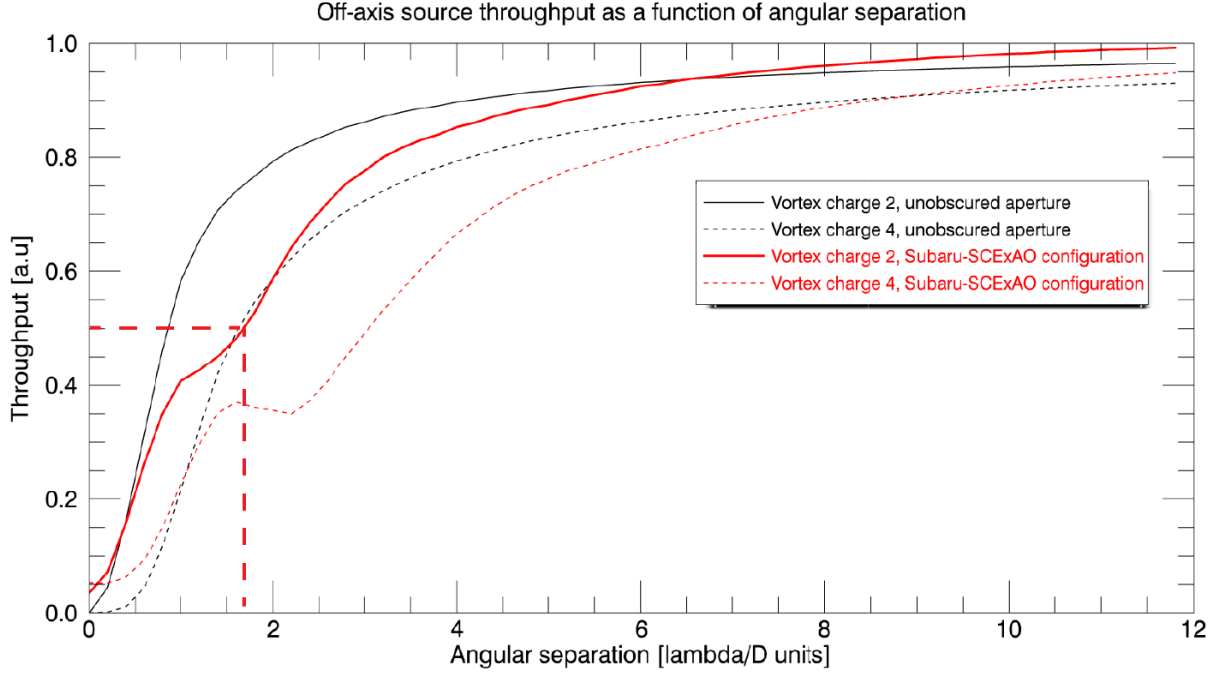

**Figure A1.** Vortex coronagraph throughput in function of angular separation for the ideal unobstructed pupil charge-2 and charge-4 cases (black curves), and the real Subaru-SCExAO configuration with a vortex chage-2 or charge-4 (red curves). The IWA (dashed red lines) of the current SCExAO charge-2 vortex is *1.7 λ/D* (70 mas).

### A.2. Coronagraphic performance metrics

It can be useful to quickly detail the few key contrast metrics usually employed to characterize a focal-plane coronagraph, as used in this work.

The coronagraphic null depth *N*, sometimes also called "leakage", is computed as the ratio of total intensity of the coronagraphic image (focal- or pupil-plane, downstream of the Lyot pupil stop) to the total intensity of non-coronagraphic image (focal- or pupil-plane, downstream, and with the same Lyot pupil stop still in place):

$$N = \frac{\int_{x,y} I_{FP}^{coro}(x,y)\,dxdy}{\int_{x,y} I_{FP}^{non-coro}(x,y)\,dxdy} = \frac{\int_{x,y} I_{PP}^{coro}(x,y)\,dxdy}{\int_{x,y} I_{PP}^{non-coro}(x,y)\,dxdy} \qquad (3)$$

The total attenuation *A*, sometimes also referred as "total rejection", is simply defined as the inverse of the null depth:

$$A = \frac{1}{N} \qquad (4)$$

The peak-to-peak attenuation $A_{PTP}$, sometimes also referred as "peak rejection", is a focal-plane metric computed as the ratio of the peak intensity in the non-coronagraphic image to the peak of the coronagraphic one:

$$A_{PTP} = \frac{\max\left[I_{FP}^{non-coro}(x,y)\right]}{\max\left[I_{FP}^{coro}(x,y)\right]} \tag{5}$$

Although a contrast curve is more useful to the observer, these metrics enable rapid evaluation and performance comparison of coronagraphic optics. In the absence of wavefront errors, $A$ and $N$ provide the approximate reduction of the PSF past a few $\lambda/D$. Those total attenuation/null depth metrics can, in general, be analytically retrieved from various leakage source contributions (see below). However they can be difficult to measure directly at the telescope, especially from single raw frames, and in the absence of basic frame cosmetics (dark subtraction, flat-fielding). Inversely, the peak-to-peak attenuation $A_{PTP}$ is much easier to evaluate directly on-sky, through a simple evaluation of peak counts and the ratio of integration times. It does also provide a direct estimation of the gain in integration time (deeper exposures in unsaturated regime) made possible by the use of the coronagraph. In the theoretical case of an unobstructed aperture, the condition $A = N^{-1} \sim A_{PTP}$ is met for small wavefront errors regime, but this quickly becomes invalid for non-ideal telescope apertures, owing to the coronagraphic PSF reshaping (e.g. donut-shaped PSF, see Fig.2). In the case of a complex telescope aperture, a constant scaling ratio between $A$ and $A_{PTP}$ can be estimated close to $SR \sim 1$ conditions. Numerical simulations with Subaru-SCExAO geometrical configuration show that this scaling factor is $A_{PTP}/A \sim 4.2$ for $SR \sim 1$ (see Fig.2). However, the presence of non-common path aberration (NCPAs) speckles interfering with the PSF, or any slight pointing error on the vortex, can drastically compromise the correct estimation of $A_{PTP}$.

*A.3. Leakage source: central obscuration*

The case of the vortex leakage due to the central obscuration (secondary mirror, see Fig.1) has already been covered in detail (Mawet et al. 2011; Serabyn et al. 2017), and the azimuthally-integrated electric field amplitude along the post-coronagraphic pupil radial separation $r$ was shown to be analytically expressed as follows for a charge-2 vortex:

$$\left|E_{before\,LS}^{leak,\sec ondary}(r)\right| = \begin{cases} 0, r < R_p \\ \left(\frac{R_s}{r}\right)^2, R_s < r < R_P \\ \left(\frac{R_p}{r}\right)^2 - \left(\frac{R_s}{r}\right)^2, r > R_P \end{cases} \tag{6}$$

where $R_P$ and $R_S$ are the primary and secondary mirror radii respectively, as in Figure 1. After the Lyot stop (LS), and given the retained undersize/oversize masking factors $\alpha$ and $\beta$ (see Table 2), we therefore get:

$$\left|E_{after\,LS}^{leak,\sec ondary}(r)\right| = \int_{\beta R_s}^{\alpha R_P} \left(\frac{R_s}{r}\right)^2 dr = R_s^2 \left(\frac{1}{\beta R_s} - \frac{1}{\alpha R_P}\right) \quad (7)$$

The corresponding null depth $N$ can then be obtained for the given Lyot stop configuration, neglecting all the effects from the spiders secondary support structure:

$$N_{\sec ondary} = \frac{I_{\sec ondary}^{leak}}{I^{non-coro}} = \left(\frac{\left|E_{PP}^{leak,\sec ondary}\right|}{\left|E_{PP}^{non-coro}\right|}\right)^2 = \left[\frac{R_S^2\left(\frac{1}{\beta R_S} - \frac{1}{\alpha R_P}\right)}{\beta R_S - \alpha R_P}\right]^2 = \left(\frac{R_S}{\alpha \beta R_P}\right)^2 \quad (8)$$

For the retained SCExAO vortex configuration presented here (see Table 2), Eq.8 gives $N_{secondary} = 4.76 \cdot 10^{-2}$, corresponding to $A \sim 21$.

*A.4. Leakage source: chromaticity*

Chromatic leakage due to the non-ideal phase profile of the vortex for various individual wavelengths (or any phase mask, with the notable exception of AGPM waveplates making use of geometric birefringence), has been shown to be expressed as (Riaud et al. 2003):

$$N_{chromatic} = \frac{\pi^2}{48 R_\lambda^2} \quad (9)$$

with $R_\lambda = \lambda_0/\Delta\lambda$ being the spectral resolution.

In the case of the H-band SCExAO vortex, with a bandwidth of ~10% around $\lambda_0 = 1.65$ μm (taking into account detector sensitivity) the chromatic null depth according to Eq. 9 is $N_{chromatic} = 2.06 \cdot 10^{-3}$, corresponding to $A \sim 486$. It is therefore a negligible source of leakage compared to the other contributors considered here.

*A.5. Leakage source: small wavefront errors*

Assuming small wavefront errors ($SR > 0.3$), Marechal's approximation for phase errors $\sigma^2 = -ln(SR)$ can be followed, and the null depth can be analytically estimated as (Boccaletti et al. 2004; Mawet et al. 2010):

$$N_{WFE} = \frac{\sigma^2}{4} = -\frac{\ln(SR)}{4} \quad (10)$$

In the extreme AO regime, and neglecting NCPAs, SCExAO being able to reach $SR \sim 0.9$ in good seeing conditions as in 2017A yields $N_{WFE} \sim 2.63 \cdot 10^{-2}$, corresponding to an attenuation $A \sim 38$. Interestingly, a Strehl ratio of ~0.96 is actually required to be able to exceed 100:1 attenuation in absence of any other contributor, illustrating the strong dependence of ground-based coronagraphy on AO and NCPA wavefront control performances.

## A.6. Leakage source: tip/tilt jitter

Tip-tilt jitter is a critical source of leakage for focal-plane coronagraphs, especially for low-IWA phase masks such as a charge-2 vortex waveplate (i.e. the gain in IWA logically means jitter has to be minimized!). Following Huby et al. 2015 derivation for off-axis transmission at small separations, Eq.1 can be used:

$$N_{jitter} \sim \frac{\pi^2 \sigma_{T-T}^2}{8} \quad (11)$$

with $\sigma_{T-T}$ being the rms tip-tilt jitter in units of $\lambda/D$.

Injecting a residual $\sigma_{T-T} \sim 0.26\ \lambda/D$ (10 mas) rms as observed in our SCExAO temporally-resolved non-coronagraphic PSF sequences of *Procyon* (see §4.1), gives $N_{jitter} \sim 8.34 \cdot 10^{-2}$, limiting the attenuation to $A \sim 12$ in absence of other contributors. This is a considerable leak, at the same order of magnitude as the leakages from the non-ideal pupil (§A.3) and residual wavefront errors for *SR ~ 0.9* (§A.5) combined! However we note that the expression of Eq. 11 is mostly valid for small tip-tilt jitter values ($\sigma_{T-T} < 0.15\ \lambda/D$ or so) and can slightly overestimate the leakage for larger jitter conditions (this is especially true for non-ideal apertures): from Huby et al. 2015 (Fig. B2) we derive a gross correction factor of ~0.67 for the true leakage around 0.25 $\lambda/D$ rms, leading to the still large analytical value of $N_{jitter} \sim 5.56 \cdot 10^{-2}$, i.e. an attenuation of $A \sim 18$ in absence of any other leakage contributors. Being able to contain tip/tilt jitter to less than 0.05 $\lambda/D$ rms would make this contribution inferior to chromatic effects (i.e. negligible).

## A.7. Combined leakage expression

In general the individual leakage terms add up incoherently (Riaud et al. 2003, Serabyn et al. 2017) to converge to the overall null depth, but here the contributions from the secondary (§A.3) and the wavefront errors (§A.5) are inherently coherent, while the other leakage terms can indeed be seen as incoherent sums of PSFs at all the wavelengths over the bandwidth, or with various tip/tilt offsets over time. Therefore we can write the following expression to combine the individual leakage contributions, assuming small phase errors [i.e. $|(E_1 + E_2)(E_1 + E_2)^*| \sim |E_1|^2 + 2|E_1||E_2| + |E_2|^2$]:

$$N_{total} = N_{coherent} + N_{chromatic} + N_{jitter} \sim \left(\sqrt{N_{secondary}} + \sqrt{N_{WFE}}\right)^2 + N_{chromatic} + N_{jitter} \quad (12)$$

Equation 12 assumes uniformly distributed wavefront phase errors across the pupil, in the sense that the term $N_{WFE}$ solely refers to atmospheric residuals and does not include contributions from non-common path aberrations (NCPAs), which are neglected in this analysis. Figure A2 plots the analytical expression of Eq.12 in function of the Strehl ratio, with various leakage terms taken into account, over the same numerically simulated plots as in Figure 3. The agreement between numerical results and analytical predictions is good up until very high Strehl ratio (*SR ~ 0.9*), beyond which the attenuation is underestimated by the analytical expression of Eq.12. This can explained by the effects of the spiders being neglected in Eq.8, when describing the leakage caused by the non-ideal pupil. Indeed, having some starlight being perfectly folded "inside" the spiders (Fig.1) – then entirely blocked by the corresponding Lyot stop spiders – enables to achieve slightly better light cancellation.

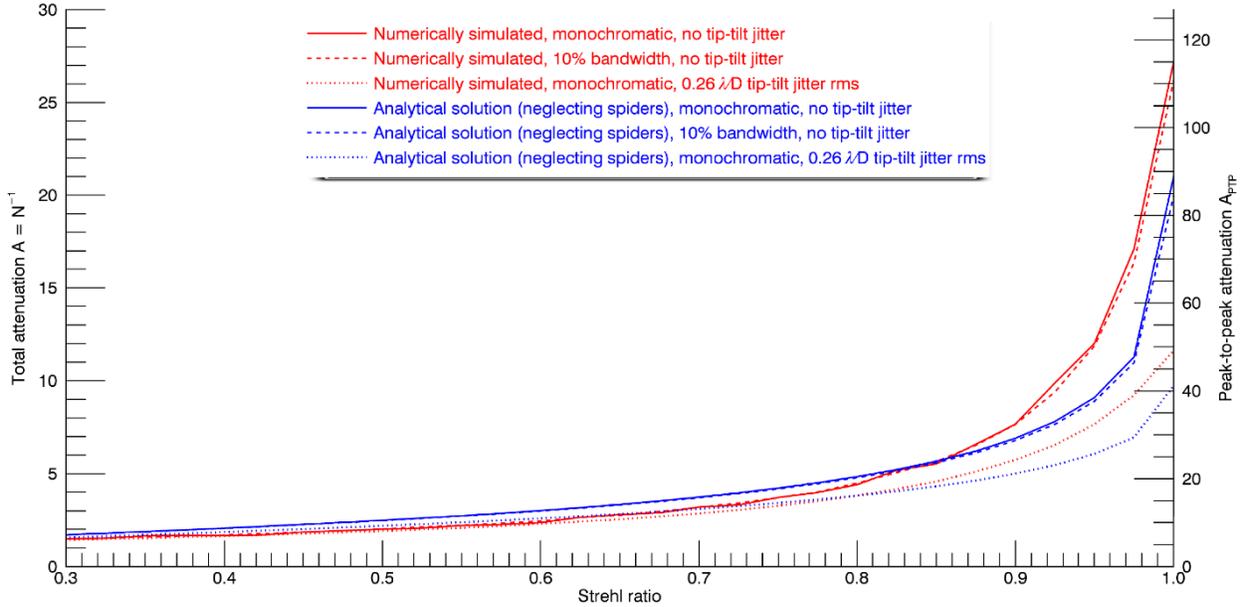

**Figure A2. Numerically simulated (red) and analytical (Eq.12, blue) SCExAO vortex null depth performance predictions in function of Strehl ratio, in the diffraction-limited regime (*SR > 0.3*). Various leakage contributors (see Annex) are added to illustrate their respective impact on null depth. The right-hand side $A_{PTP}$ vertical axis assumes a coronagraphic PSF geometric scaling factor $A_{PTP}/A \sim 4.2$, but this quickly loses validity for *SR < 1*.**

## REFERENCES

Amara, A., & Quanz, S. P. 2012, MNRAS, 427, 948

Beuzit, J.-L., Feldt, M., Dohlen, K. 2008, Proc. SPIE, 7014, 701418

Boccaletti, A., Riaud, P., Baudoz, P., Baudrand, J., Rouan, D., Gratadour, D., Lacombe, F., & Lagrange, A.-M. 2004, PASP, 116, 825

Bottom, M., Kuhn, J., Mennesson, B., Mawet, D., Shelton, J. C., Wallace, J. K., & Serabyn, E. 2015, ApJ, 809, 1

Bottom, M., Shelton, J. C., Wallace, J. K., Bartos, R., Kuhn, J., Mawet, D., Mennesson, B., Burruss, R., et al. 2016, PASP, 128, 965

Bottom, M., Wallace, J. K., Bartos, R. D., Shelton, J. C., & Serabyn, E. 2017, MNRAS, 464, 3

Bowler, B. P. 2016, PASP, 128, 102001

Carson, J., Thalmann, C., Janson, M., Kozakis, T., Bonnefoy, M., Biller, B., Schlieder, J., Currie, T., et al. 2013, ApJL, 763, 2

Currie, T., Debes, J., Rodigas, T. J., Burrows, A., Itoh, Y., Fukagawa, M., Kenyon, S. J., Kuchner, M., et al. 2012, ApJL, 760, 2

Currie, T., Guyon, O., Tamura, M., Kudo, T., Jovanovic, N., Lozi, J., Schlieder, J. E., Brandt, T. D., et al. 2017, ApJ, 836, 1

Delacroix, C., Absil, O., Forsberg, P., Mawet, D., Christiaens, V., Karlsson, M., Boccaletti, A., Baudoz, P., et al. 2013, A&A, 533, A98
Gomez Gonzales, C. A., Absil, O., Absil, P.-A., Van Droogenbroeck, M., Mawet, D., & Surdej, J. 2016, A&A, 589, A54
Gomez Gonzalez, C. A., Wertz, O., Absil, O., Christiaens, V., Defrère, D., Mawet, D., Milli, J., Absil, P.-A., et al. 2017, AJ, 154, 1
Groff, T. D., Kasdin, N. J., Limbach, M. A., Galvin, M., Carr, M. A., Knapp, G., Brandt, T., Loomis, C., et al. 2015, Proc. of SPIE, 9605, 96051C
Guyon, O., Pluzhnik, E. A., Galicher, R., Martinache, F., Ridgway, S. T., Woodruff, R. A. 2005, ApJ, 662, 744
Guyon, O., Martinache, F., Belikov, R., & Soummer, R. 2010, ApJS, 190, 220
Hodapp, K. W., Suzuki, R., Tamura, M., Abe, L., Suto, H., Kandori, R., Morino, J., Nishimura, T., et al. 2008, Proc. of SPIE, 7014, 701419
Huby, E., Perrin, G., Marchis, F., Lacour, S., Kotani, T., Duchêne, G., Choquet, E., Gates, E. L., et al. 2012, A&A, 541, A55
Huby, E., Baudoz, P., Mawet, D., & Absil, O. 2015, A&A, 584, A74
Jenkins, C. 2008, MNRAS, 384, 515
Jones, J., White, R. J., Quinn, S., Ireland, M., Boyajian, T., Schaefer, G., & Baines, E. K. 2016, ApJL, 822, 1
Jovanovic, N., Martinache, F., Guyon, O., Clergeon, C., Singh, G., Kudo, T., Garrel, V., Newman, K., et al. 2015a, PASP, 127, 890
Jovanovic, N., Guyon, O., Martinache, F., Pathak, P., Hagelberg, J., & Kudo, T. 2015b, ApJL, 813, 2
Kenworthy, M. A., Codona, J. L., Hinz, P. M., Angel, J. R. P., Heinze, A., & Sivanandam, S. 2007, ApJ, 660, 762
Lafrenière, D., Doyon, R., Nadeau, D., Artigau, É., Marois, C., & Beaulieu, M. 2007a, ApJ, 661, 2
Lafrenière, D., Marois, C., Doyon, R., Nadeau, D., & Artigau, É. 2007b, ApJ, 660, 770
Lozi, J., Guyon, O., Jovanovic, N., Singh, G., Goebel, S., Norris, B., & Okita, H. 2016, Proc. of SPIE, 9909, 99090J
Macintosh, B. A., Anthony, A., Atwood, J. et al. 2012, Proc. SPIE, 8446, 84461U
Marois, C., Lafreniere, D., Doyon, R., Macintosh, B., & Nadeau, D. 2006, ApJ, 641, 556
Marois, C., Macintosh, B., & Veran, J.-P. 2010, Proc. of SPIE, 7736, 77361J
Marsden, D., Mazin, B. A., Bumble, B., Meeker, S., O'Brien, K., McHugh, S., Strader, M., & Langman, E. 2012, Proc. of SPIE, 8453, 84530B
Martinache, F., Guyon, O., Jovanovic, N., Clergeon, C., Singh, G., & Kudo, T. 2014, Proc. of SPIE, 9148, 914821
Martinache, F., Jovanovic, N., & Guyon, O. 2016, A&A, 593, A33
Mawet, D., Serabyn, E., Liewer, K., Hanot, C., McEldowney, S., Shemo, D., & O'Brien, N. 2009, Optics Express, 17, 1902
Mawet, D., Serabyn, E., Liewer, K., Burruss, R., Hickey, J., & Shemo, D. 2010, ApJ, 709, 53
Mawet, D., Serabyn, E., Wallace, J. K., & Pueyo L. 2011, Optics Letters, 36, 8
Mawet, D., Absil, O., Delacroix, C., Girard, J. H., Milli, J., O'Neal, J., Baudoz, P., Boccaletti, A., et al. 2013a, A&A, 552, L13
Mawet, D., Pueyo, L., Carlotti, A., Mennesson, B., Serabyn, E., & Wallace, J. K. 2013b, ApJS, 209, 1

Mawet, D., Milli, J., Wahhaj, Z., Pelat, D., Absil, O., Delacroix, C., Boccaletti, A., Kasper, M., et al. 2014, ApJ, 792, 2
McEldowney, S. C., Shemo, D. M., Chipman, R. A., & Smith, P. K. 2008, Optics Letters, 33, 2
Minowa, Y., Hayano, Y., Oya, S., Watanabe, M., Hattori, M., Guyon, O., Egner, S., Saito, Y., et al. 2010, Proc. of SPIE, 7736, 77363N
Murakami, N., Guyon, O., Martinache, F., Matsuo, T., Yokochi, K., Nishikawa, J., Tamura, M., Kurokawa, N., et al. 2010, Proc. SPIE, 7735, 773533
N'Diaye, M., Vigan A., Dohlen, K., Sauvage, J.-F., Caillat, A., Costille, A., Girard, J. H. V., Beuzit, J.-L. et al. 2016, A&A, 592, A79
Nersisyan, S. R., Tabiryan, N, V., Mawet, D, & Serabyn, E. 2013, Optics Express, 21, 7
Norris, B., Schworer, G., Tuthill, P., Jovanovic, N., Guyon, O., Stewart, P., & Martinache, F. 2015, MNRAS, 447, 3
Riaud, P., Boccaletti, A., Baudrand, J., & Rouan D. 2003, PASP, 115, 712
Rouan, D., Riaud, P., Boccaletti, A., Clénet, Y., & Labeyrie A. 2000, PASP, 112, 1479
Schmid, H. M., Bazzon, A., Milli, J., Roelfsema, R., Engler, N., Mouillet, D., Lagadec, E., Sissa, E., et al. 2017, A&A, 602, A53
Serabyn, E., Mawet, D., & Burruss, R. 2010, Nature, 464, 7291
Serabyn, E., Huby, E., Matthews, K., Mawet, D., Absil, O., Femenia, B., Wizinowich, P., Karlsson, M., et al. 2017, AJ, 153, 1
Singh, G., Lozi, J., Guyon, O., Baudoz, P., Jovanovic, N., Martinache, F., Kudo, T., Serabyn, E., et al. 2015, PASP, 127, 955
Soummer, R., Pueyo, L. & Larkin, J. 2012, ApJL, 755, L28
Uyama, T., Hashimoto, J., Kuzuhara, M., Mayama, S., Akiyama, E., Currie, T., Livingston, J., Kudo, T., & al. 2017, AJ, 153, 3
Wallace, J. K., Burruss, R. S., Bartos, R. D., Trinh, T. Q., Pueyo, L. A., Fregoso, S. F., Angione, J. R., & Shelton, J. C. 2010, Proc. SPIE, 7736, 77365D